\documentclass[prc,aps,floatfix,twocolumn,groupedaddress,nofootinbib,showpacs,preprintnumbers,
amsmath,amssymb,amsfonts,widetable] {revtex4-1}
\usepackage{bm}
\usepackage{soul}
\usepackage{array}
\usepackage{lipsum}
\usepackage{relsize}
\usepackage{mathrsfs}
\usepackage{amssymb}
\usepackage{amsmath}
\usepackage{graphicx}
\usepackage[usenames]{color}
\usepackage{pslatex}
\usepackage{booktabs}
\usepackage{physics}

\newcommand{\Edens}{{\mathlarger{\varepsilon}}}
\newcommand{\EoverA}{{\mathlarger{\epsilon}}}
\newcommand{\chisq}{{\chi}^{\raisebox{-1pt}{$\scriptstyle{2}$}}}

\begin{document}
\title{Bayesian refinement of covariant energy density functionals} 

\author{Marc Salinas and J. Piekarewicz}

\affiliation{Department of Physics, Florida State University, 
Tallahassee, FL 32306, USA}

\date{\today}

\begin{abstract}
The last five years have seen remarkable progress in our quest to determine the equation of 
state of neutron rich matter. Recent advances across the theoretical, experimental, and 
observational landscape have been incorporated in a Bayesian framework to refine existing 
covariant energy density functionals previously calibrated by the properties of finite nuclei. 
In particular, constraints on the maximum neutron star mass from pulsar timing, on stellar
radii from the NICER mission, on tidal deformabilities from the LIGO-Virgo collaboration, 
and on the dynamics of pure neutron matter as predicted from chiral effective field theories, 
have resulted in significant refinements to the models, particularly to those predicting a stiff 
symmetry energy. Still, even after these improvements, we find challenging to reproduce 
simultaneously the neutron skin thickness of both ${}^{208}$Pb and ${}^{48}$Ca recently
reported by the PREX/CREX collaboration.
\end{abstract}

\maketitle

\section{Introduction}
We have entered the golden era of neutron stars\,\cite{Baym:2019,Baym:2017whm}.  A confluence 
of pioneering discoveries during the last five years have provided stringent constraints on the equation 
of state (EOS) of neutron star matter over an enormous range of densities. Below nuclear matter 
saturation density $\rho_{0}\!\approx\!0.15\,{\rm fm}^{-3}$---a region that is difficult to probe in 
laboratory experiments---theoretical predictions of the EOS of pure neutron matter based on 
chiral effective field theory are providing valuable insights\,\cite{Hebeler:2009iv,Tews:2012fj,
Kruger:2013kua,Lonardoni:2019ypg,Drischler:2021kxf,Sammarruca:2021mhv,
Sammarruca:2022ser}. In the laboratory, the Lead Radius Experiment (PREX) 
has established that the neutron skin of ${}^{208}$Pb is thick\,\cite{Abrahamyan:2012gp,
Horowitz:2012tj,Adhikari:2021phr} suggesting, in turn, that the EOS of neutron rich matter in the 
vicinity of saturation density is stiff\,\cite{Reed:2021nqk}. Note that a ``stiff'' equation of state is 
one in which the pressure increases rapidly with increasing density, whereas one in which the 
pressure increases slowly is ``soft". In this manner, chiral EFT and laboratory experiments define 
the first rung in a ``density ladder" consisting of theoretical, experimental, and observational rungs 
that inform the EOS in a suitable density regime\,\cite{Piekarewicz:2022ycz}. In turn, neutron star 
radii are most sensitive to the EOS in the neighborhood of twice nuclear matter saturation. As such, 
stellar radii inferred from both the tidal deformability of GW170817\,\cite{Abbott:PRL2017,Abbott:2018exr} 
and from monitoring stellar hot spots by the Neutron Star Interior Composition Explorer 
(NICER)\,\cite{Riley:2019yda,Miller:2019cac,Miller:2021qha,Riley:2021pdl}, are providing the 
most sensitive constraints on the EOS at about $2\rho_{0}$. Finally, the most stringent constraints on 
the EOS at the highest densities encountered in the stellar core are obtained from the identification 
of neutron stars with masses in the in the vicinity of two solar 
masses\,\cite{Demorest:2010bx,Antoniadis:2013pzd,Cromartie:2019kug,Fonseca:2021wxt}. 

While all these discoveries are painting a fairly consistent and compelling picture of the EOS, there 
are a few instances that suggest a possible tension. First, chiral EFT calculations tend to predict a 
softer EOS at saturation density\,\cite{Drischler:2021kxf} as compared to an analysis based on the 
PREX measurement, which instead suggests a fairly stiff equation of state\,\cite{Reed:2021nqk}. 
Given that the PREX error bars are relatively large\,\cite{Adhikari:2021phr}, an improved experiment 
(``MREX") planned at the future MESA facility in Mainz should be able to confirm whether the 
discrepancy is real. Second, most nuclear structure models find a strong correlation between 
the thickness of the neutron skin in ${}^{48}$Ca and ${}^{208}$Pb. However, this correlation
is in stark disagreement with experiment. The CREX collaboration has recently reported a 
neutron skin thickness in ${}^{48}$Ca\,\cite{Adhikari:2022kgg} that is significantly smaller than 
any theoretical prediction that reproduces the large value of the neutron skin in ${}^{208}$Pb. 
Finally, the historic detection of gravitational waves by the LIGO-Virgo collaboration from the
binary neutron star coalescence GW170817 suggests that the EOS is soft in the vicinity of 
$2\rho_{0}$\,\cite{Abbott:2018exr}, although a reanalysis\,\cite{Gamba:2020wgg} could 
accommodate a stiffer EOS that brings the radius of a $1.4\,{\rm M}_{\odot}$ neutron star 
into agreement with the NICER results\,\cite{Riley:2019yda,Miller:2019cac} and with other 
constraints obtained from the analysis of the electromagnetic counterpart\,\cite{Radice:2018ozg}. 

Although the next few years will be instrumental in resolving these possible discrepancies, as of
today this situation suggests an intriguing possibility. So far, we have learned that the equation 
of state evolves from stiff at typical nuclear densities, to soft at slightly higher densities, and ultimately 
back to stiff at the highest densities encountered in the core of massive neutron stars. If confirmed, 
such an evolution from stiff, to soft, and back to stiff may be suggestive of a phase transition in the 
stellar interior.

Inspired by a recent approach that incorporates predictions from chiral effective field theory
($\chi$EFT)\,\cite{Alford:2022bpp}, we aim to refine existing covariant energy density 
functionals (EDFs)---calibrated to the ground state properties of finite nuclei---by incorporating 
both $\chi$EFT predictions for the EOS of pure neutron matter together with observational 
constraints provided by LIGO-Virgo, NICER, and pulsar timing. The implementation of this new 
calibration procedure uses a covariance matrix collected from existing EDFs\,\cite{Chen:2014sca,
Chen:2014mza} as the prior distribution of parameters. This prior distribution is then combined 
with a likelihood function that incorporates all the new information. It is the primary goal of this 
paper to refine existing EDFs by incorporating both theoretical and observational constraints in 
a consistent Bayesian framework. 

The manuscript has been organized as follows. In Sec.\ref{Sec:Formalism} we define the
structure of the covariant EDF that we aim to refine. In the same section we discuss our
choice of posterior distribution composed from a prior distribution obtained from a previous 
parameter estimation and a likelihood function informed by the new data. In Sec.\ref{Sec:Results} 
we analyze the impact of the new data on a particular set of isovector sensitive observables. 
One of the goals of the present study is to determine whether the wealth of new information 
incorporated in the calibration demands an extension of the relatively simple isovector sector 
of the kind of EDF used in this work. Finally, in Sec.\ref{Sec:Conclusions} we summarize our 
results and provide an outlook on how to improve the synergy between nuclear physics and 
observational astronomy.

\section{Formalism}
\label{Sec:Formalism}
\subsection{Covariant Density Functional Theory}

In the framework of covariant density functional theory (DFT), the underlying degrees of freedom 
are nucleons interacting via the exchange of three mesons and the photon. In the particular version 
of covariant DFT that will be employed here, the interactions are encoded in an effective Lagrangian 
density containing conventional Yukawa couplings plus meson self-interactions:\,\cite{Walecka:1974qa,
Boguta:1977xi,Serot:1984ey,Mueller:1996pm,Serot:1997xg,Horowitz:2000xj}:
\begin{widetext}
\begin{eqnarray}
{\mathscr L}_{\rm int} =
\bar\psi \left[g_{\rm s}\phi   \!-\! 
         \left(g_{\rm v}V_\mu  \!+\!
    \frac{g_{\rho}}{2}{\mbox{\boldmath $\tau$}}\cdot{\bf b}_{\mu} 
                               \!+\!    
    \frac{e}{2}(1\!+\!\tau_{3})A_{\mu}\right)\gamma^{\mu}
         \right]\psi \!-\!
    \frac{\kappa}{3!} (g_{\rm s}\phi)^3 \!-\!
    \frac{\lambda}{4!}(g_{\rm s}\phi)^4 \!+\!
    \frac{\zeta}{4!}   g_{\rm v}^4(V_{\mu}V^\mu)^2 +
   \Lambda_{\rm v}\Big(g_{\rho}^{2}\,{\bf b}_{\mu}\cdot{\bf b}^{\mu}\Big)
                           \Big(g_{\rm v}^{2}V_{\nu}V^{\nu}\Big)\;.
 \label{LDensity}
\end{eqnarray}
\end{widetext}

Here $\psi$ is the isodoublet nucleon field, $A_{\mu}$ is the photon field, and 
$\phi$, $V_{\mu}$, and ${\bf b}_{\mu}$ represent the isoscalar-scalar 
$\sigma$-meson, the isoscalar-vector $\omega$-meson, and the isovector-vector 
$\rho$-meson fields, respectively. The $\sigma$-meson is responsible for the 
intermediate range attraction of the nuclear force, the $\omega$-meson mediates 
the repulsion at short distances, while the $\rho$-meson provides a dynamical
contribution to the nuclear symmetry energy. We note that the isovector sector 
of the model is entirely defined in terms of two model parameters: the Yukawa 
coupling constant $g_{\rho}$ and the mixed isoscalar-isovector coupling 
$\Lambda_{\rm v}$, introduced to modify the density dependence of the symmetry 
energy\,\cite{Horowitz:2000xj}. Although in the spirit of an effective field theory one 
should incorporate all possible meson interactions that are allowed by symmetry
considerations to a given order in a power-counting scheme, until recently the data 
base of isovector observables was too limited to justify the inclusion of additional 
parameters. However, the wealth of new experimental and observational data 
collected within the last few years may demand extensions to the isovector sector
of the model. 

\subsection{Mean Field Approximation}
\label{Sec:MFA}

In the study of uniform neutron rich matter, the field equations resulting from the above 
Lagrangian density may be solved exactly in the mean field approximation. Assuming a 
static and uniform ground state, the meson fields may be replaced by their classical,
ground-state expectation values\,\cite{Walecka:1974qa,Serot:1984ey}:
\begin{subequations}
\begin{align}
\phi  \rightarrow & \langle\phi\rangle=\phi_{0},\\
V^{\mu}  \rightarrow & \langle V^{\mu}\rangle=g^{\mu0}V_{0},\\
b_{a}^{\mu}  \rightarrow & \langle b_{a}^{\mu}\rangle=g^{\mu0}\delta_{a3}b_{0},
\end{align}
\end{subequations}
which in turn satisfy the following mean field equations:
\begin{subequations}
\begin{align}
  &  \frac{m_{\rm s}^{2}}{g_{\rm s}^{2}}\Phi_{0}+\frac{\kappa}{2}\Phi_{0}^{2}
  + \frac{\lambda}{6}\Phi_{0}^{3} = \big(\rho_{\rm sp}\!+\!\rho_{\rm sn}\big), \\
 &  \frac{m_{\rm v}^{2}}{g_{\rm v}^{2}}W_{0} + \frac{\zeta}{6}W_{0}^{3} 
  + 2\Lambda_{\rm v}B_{0}^{2}W_{0} = \big(\rho_{\rm vp}\!+\!\rho_{\rm vn}\big), \\
&  \frac{m_{\rho}^{2}}{g_{\rho}^{2}}B_{0} 
  + 2\Lambda_{\rm v}W_{0}^{2}B_{0} = \frac{1}{2}\big(\rho_{\rm vp}\!-\!\rho_{\rm vn}\big).
\end{align}
\label{KGEqns}
\end{subequations} 
Here the source terms are written in terms of scalar and time-like vector densities for
both protons and neutrons. Moreover, we have defined $\Phi_{0}\!\equiv\!g_{\rm s}\phi_{0}$,
$W_{0}\!\equiv\!g_{\rm v}V_{0}$, and $B_{0}\!\equiv\!g_{\rho}b_{0}$. Note that the 
Klein-Gordon equations for the meson fields depend on the ratio of the coupling constant 
to the corresponding meson mass. Such a degeneracy in model parameters can only be 
broken by invoking finite-nucleus observables.

In turn, the nucleon satisfies a Dirac equation with an effective mass $M^{\star}\!\equiv\!M\!-\!\Phi_{0}$ 
and a dispersion relation given by
\begin{equation}
 {\mathlarger{\varepsilon}}_{p,n}({\bf k})=\sqrt{k^{2}+M^{\star\,2}}+W_{0}\pm\frac{1}{2}B_{0},
\end{equation}
where $M$ is the free nucleon mass and ${\bf k}$ is the nucleon momentum.

\subsection{Equation of State}
\label{Sec:EoS}

At zero temperature and for a given energy density, the EOS of neutron star matter is obtained by 
computing the associated pressure provided by a charge-neutral system of nucleons and leptons 
in beta equilibrium. Although such is the EOS that appears in the Tolman-Oppenheimer-Volkoff 
(TOV) equations, it is instructive to start with the EOS of infinite nuclear matter---an idealized system 
of protons and neutrons interacting solely via the strong nuclear force. In this limit, the electroweak 
sector is effectively turned off, so both proton and neutron densities are individually conserved. 
As such, the EOS of infinite nuclear matter in the mean-field approximation may be written as
\begin{widetext}
\begin{equation}
    \Edens_{\rm nuc}(\rho_{p},\rho_{n}) =  \Edens_{p}(\rho_{p}) + \Edens_{n}(\rho_{n}) 
    +  \left(\frac{m_{\rm s}^{2}}{2g_{\rm s}^{2}}\Phi_{0}^{2} + 
  \frac{\kappa}{6}\Phi_{0}^{3} + \frac{\lambda}{24}\Phi_{0}^{4}\right) 
   + \left(\frac{m_{\rm v}^{2}}{2g_{\rm v}^{2}}W_{0}^{2} + \frac{\zeta}{8}W_{0}^{4} 
     +        \frac{m_{\rho}^{2}}{2g_{\rho}^{2}}B_{0}^{2} + 3\Lambda_{\rm v}W_{0}^{2}B_{0}^{2}\right),
\label{EDensity0}
\end{equation}
\end{widetext}
where $\Edens_{p}(\rho_{p})$ and $\Edens_{n}(\rho_{n})$ are energy densities of a free Fermi 
gas of particles of mass $M^{\star}$ and density $\rho_{p}$ and $\rho_{n}$, respectively. In turn, 
the associated pressure can be obtained by invoking the Hugenholtz-Van Hove theorem for 
asymmetric nuclear matter\,\cite{Satpathy:1983}. That is,
\begin{equation}
 \Edens_{\rm nuc}(\rho_{p},\rho_{n})  \!+\! P_{\rm nuc}(\rho_{p},\rho_{n}) \!=\! 
 \rho_{p}E_{{\rm F}p}+\rho_{n}E_{{\rm F}n},
 \label{HVH}
\end{equation}
where the proton and neutron Fermi energies are 
\begin{subequations}
\begin{align}
 E_{{\rm F}p} & = \sqrt{k_{\rm Fp}^{2}+M^{\star 2}}+\big(W_{0}+\frac{1}{2}B_{0}\big), \\
 E_{{\rm F}n} & = \sqrt{k_{\rm Fn}^{2}+M^{\star 2}}+\big(W_{0}-\frac{1}{2}B_{0}\big).
\end{align}
\label{EFermi}
\end{subequations}
Here the Fermi momenta are related to the corresponding densities as follows:
\begin{equation}
 \rho_{p} = \frac{k_{\rm Fp}^{3}}{3\pi^{2}}
  \hspace{5pt} {\rm and} \hspace{5pt}
 \rho_{n} = \frac{k_{\rm Fn}^{3}}{3\pi^{2}}.
\label{kFermi}
\end{equation}
Finally, the energy density of a free Fermi gas of particles of mass $M$ and density 
$\rho\!=\!k_{\rm F}^{3}/3\pi^{2}$ may be computed in closed form:
\begin{align}
 \Edens(\rho) & = \frac{1}{\pi^{2}} \int_{0}^{k_{\rm F}} k^{2}\sqrt{k^{2}+M^{2}}\,dk \nonumber \\
    & = \frac{M^{4}}{8\pi^{2}}\Big[ x_{\rm F}y_{\rm F}\big(x_{\rm F}^{2}+y_{\rm F}^{2}\big)
        -\ln\big(x_{\rm F}+y_{\rm F}\big)\Big],
\label{EDensity1}
\end{align}
where $x_{\rm F}\!=\!k_{\rm F}/M$ and $y_{\rm F}\!=\!\sqrt{1+x_{\rm F}^{2}}$. This expression is 
useful in the evaluation of the proton and neutron energy density appearing in Eq.(\ref{EDensity0})
as well as in computing the energy density of the leptonic contribution to the EOS of neutron star 
matter. 

Indeed, the energy density of neutron star matter may be written as a sum of the nuclear
contribution [Eq.(\ref{EDensity0})] plus a leptonic contribution involving both electrons and
muons. That is, 
\begin{equation}
  \Edens(\rho,Y_{p},Y_{e}) = \Edens_{\rm nuc}(\rho_{p},\rho_{n}) 
                            + \Edens_{\rm e}(\rho_{e}) + \Edens_{\rm \mu}(\rho_{\mu}),
\label{ErhoYpYe}
\end{equation}
where the conserved baryon (or vector) density is $\rho\!=\rho_{n}\!+\!\rho_{p}$ and the individual 
nucleonic and leptonic densities are defined in terms of suitable proton and electron fractions 
$Y_{p}$ and $Y_{e}$ as follows:
\begin{subequations}
\begin{align}
  & \frac{\rho_{p}}{\rho}=Y_{p} \hspace{8pt}{\rm and}\hspace{8pt} \frac{\rho_{n}}{\rho} = 1-Y_{p}, \\
  & \frac{\rho_{e}}{\rho}=Y_{e} \hspace{8pt}{\rm and}\hspace{8pt} \frac{\rho_{\mu}}{\rho} = Y_{p}-Y_{e}.
\label{Rhos}
\end{align}
\end{subequations}

Given that the matter inside neutron stars is fully catalyzed, both the proton and electron 
fractions adjust themselves through weak interactions to reach the absolute ground state at a
given baryon density $\rho$. Hence, $Y_{p}$ and $Y_{e}$ are determined by demanding that 
\begin{equation}
 \left(\frac{\partial\Edens(\rho,Y_{p},Y_{e})}{\partial Y_{p}}\right)_{\!\!\rho,Y_{e}} \!\!=
 \left(\frac{\partial\Edens(\rho,Y_{p},Y_{e})}{\partial Y_{e}}\right)_{\!\!\rho,Y_{p}} \!\!= 0.
\label{dEdxdy}
\end{equation}
These conditions are entirely equivalent to demanding beta equilibrium through the following 
reactions:
\begin{subequations}
\begin{align}
  & n \leftrightarrow p + e^{-} + \bar{\nu}_{e}  \Rightarrow \mu_{n}=\mu_{p}+\mu_{e}, \\
  & \mu^{-} \leftrightarrow e^{-} + \bar{\nu}_{e} + \nu_{\mu} \Rightarrow \mu_{\mu}=\mu_{e},
\label{BetaEq}
\end{align}
\end{subequations}
where $\mu_{x}$ is the chemical potential of the various species and the neutrino chemical 
potential has been neglected.

Besides the uniform stellar core, a neutron star contains a solid crust that develops once the uniform 
ground state becomes unstable against clustering correlations. Given the short-range nature of the
nuclear force, it becomes energetically favorable for nucleons to cluster as soon as the average
inter-particle separation becomes larger than the range of the nucleon-nucleon interaction. At the
low densities found in the outer stellar crust, the system forms a Coulomb lattice of neutron-rich 
nuclei embedded in a degenerate electron gas\,\cite{Baym:1971pw, RocaMaza:2008ja}. In this
region, the pressure support against gravitational collapse is provided by the degenerate electrons.  
Hence, the EOS for this region is relatively well known\,\cite{Feynman:1949zz,Baym:1971pw,Haensel:1989}.
However, at a density of about $2.6\!\times\!10^{-4}\,{\rm fm}^{-3}$, the nuclei in the outer crust 
become so neutron rich that no more neutrons can be bound. Such ``neutron-drip" region delineates
the boundary between the outer and the inner crust. The inner stellar crust extends from the 
neutron-drip density up to about $\rho\!\approx\!2/3 \rho_0$, where the uniformity in the system 
is restored. The precise value of the crust-core transition density is unknown as it depends on the 
stiffness of the EOS of neutron rich matter below saturation density. Besides the formation of a Coulomb 
crystal of neutron-rich nuclei embedded in a uniform electron gas and a dilute superfluid 
neutron gas\,\cite{Piekarewicz:2013dka}, the inner crust exhibits complex and exotic structures 
that are collectively known as``nuclear pasta"\,\cite{Ravenhall:1983uh,Hashimoto:1984,Lorenz:1992zz}.
The complex dynamics in this region is important for the understanding of transport properties as 
well as for the interpretation of cooling observations\,\cite{Fattoyev:2017ybi}. Yet, their impact on
the EOS is minimal, so for this region we resort to the equation of state of  Negele and 
Vautherin\,\cite{Negele:1971vb}. 

All that remains is to compute the transition density from the solid crust to the uniform liquid core. To 
do so, we improve on an earlier study that determines the transition density by examining the stability 
of the ground state against small density perturbations\,\cite{Carriere:2002bx} to one that relies on the 
Thermodynamic Stability Method outlined by Kubis in Ref.\,\cite{Kubis:2006kb}. From 
thermodynamic first principles the stability of the system requires the following conditions to hold true:
\begin{subequations}
\begin{align}
 & -\left(\frac{\partial P}{\partial v}\right)_{\!\!q}\!\!>0 
  \hspace{8pt}{\rm and}\hspace{8pt}
-\left(\frac{\partial \mu}{\partial q}\right)_{\!\!P}\!\!>0, 
 \label{stabilityA} \\ 
 & {\rm or} \nonumber \\
 & -\left(\frac{\partial P}{\partial v}\right)_{\!\!\mu}\!\!>0 
  \hspace{8pt}{\rm and}\hspace{8pt}
-\left(\frac{\partial \mu}{\partial q}\right)_{\!\!v}\!\!>0, 
\label{stabilityB} 
\end{align}
\label{stability}
\end{subequations}
where $v\!=\!V/A$ and $q\!=\!Q/A$ are the volume and charge per baryon, respectively---and it
has been argued in Ref.\,\cite{Kubis:2006kb} that both pair of inequalities are equivalent. The 
first inequality in Eq.(\ref{stabilityA}) ensures that the thermal incompressibility remains positive 
whereas the second one embodies the stability of charge fluctuations. Moreover, it has been
shown that for a symmetry energy that remains positive in the region of interest, the second
condition in Eq.(\ref{stabilityB}) is always satisfied\,\cite{Routray:2016yvp}. Finally, the first 
inequality in Eq.(\ref{stabilityB}) can be recasted into a more useful form that will be used here 
to compute the crust-core transition density. That is, 
 \begin{equation} \label{stability_E}
  2\rho\bigg{(} \dfrac{\partial\EoverA}{\partial\rho} \bigg{)} + 
  \rho^2 \bigg{(} \dfrac{\partial^2\EoverA}{\partial\rho^2} \bigg{)} - 
  \bigg{(} \rho \dfrac{\partial^2\EoverA}{\partial \rho\,\partial Y_p} \bigg{)}^2 \!\bigg{/}\!
  \bigg{(} \dfrac{\partial^2\EoverA}{\partial Y_p^2} \bigg{)} \!>\! 0.
\end{equation}
For a given EOS, the transition density occurs when the above inequality is violated. Although the 
parabolic approximation has been used to compute the transition density, Routray et al., have 
shown that the transition density may be overestimated by about 25 percent\,\cite{Routray:2016yvp}.
Thus, we compute the transition density without approximations by solving Eq.(\ref{stability_E}) after 
re-expressing the above inequality in a more convenient form that only involves the nucleonic chemical 
potentials, as in Eq.(8) on Ref.\,\cite{Routray:2016yvp}. Finally, we note that the thermodynamic stability 
method yields slightly lower transition densities than the dynamical method based on the Random Phase 
Approximation\,\cite{Carriere:2002bx}.

\subsection{The Nuclear Symmetry Energy}
\label{Sec:SymmE}

In this section we introduce a critical component of the energy of neutron rich
matter: the nuclear symmetry energy.  To do so, we express the energy per nucleon in 
terms of the total baryon density $\rho$ and the neutron-proton asymmetry 
$\alpha\!\equiv\!(\rho_{n}\!-\!\rho_{p})/(\rho_{n}\!+\!\rho_{p})$. Moreover, since the 
neutron-proton asymmetry is constrained to the interval $|\alpha|\!\le\!1$, the total 
energy per particle is customarily expanded in a power series in $\alpha^{2}$. That is,
\begin{equation}
  \frac{E}{A}(\rho,\alpha) -\!M = 
   \EoverA_{{}_{\rm SNM}}(\rho)
   + \alpha^{2}{\cal S}(\rho) + {\cal O}(\alpha^{4}) \,.
 \label{EoverA}
\end {equation}
The leading term in this expansion is independent of $\alpha$ and represents the energy per nucleon 
of symmetric nuclear matter. In turn, the first-order correction to the symmetric limit is the nuclear
symmetry energy ${\cal S}(\rho)$, which quantifies the energy cost in turning protons into neutrons 
(or viceversa). Note that no odd powers of $\alpha$ appear as the nuclear force is assumed to be 
isospin symmetric: in the absence of electroweak interactions it is equally costly to turn protons into 
neutrons than neutrons into protons. In the often-used ``parabolic" approximation in which
all corrections beyond second order in $\alpha$ are neglected, the symmetry energy quantifies the 
energy cost in turning symmetric nuclear matter into pure neutron matter. Finally, the behavior of
neutron rich matter in the vicinity of saturation density can be characterized in terms of a few bulk 
parameters as follows\,\cite{Piekarewicz:2008nh}:
\begin{subequations}
\begin{align}
 & \EoverA_{{}_{\rm SNM}}(\rho) = 
    \EoverA_{0} + \frac{1}{2}K_{0}x^{2} +\ldots \\
 & {\cal S}(\rho) = J + Lx + \frac{1}{2}K_{\rm sym}x^{2}+\ldots   
\end{align} 
\label{EandS}
\end{subequations}
where $x\!=\!(\rho-\rho_{0})/3\rho_{0}$ is a dimensionless parameter that quantifies the deviations 
of the density from its value at saturation. Here $\EoverA_{0}$ and $K_{0}$ are the binding energy
per nucleon and incompressibility coefficient of symmetric nuclear matter at saturation density; $J$ 
and $K_{\rm sym}$ are the corresponding terms in the symmetry energy. Note that no linear term 
in $x$ appears in $\EoverA_{{}_{\rm SNM}}$ because symmetric nuclear matter saturates, namely, 
the pressure at saturation density vanishes. Such a linear term is no longer absent from the symmetry
energy. Rather, the slope of the symmetry energy $L$ is a critical parameter that encapsulates the 
stiffness of the equation of state at saturation density. Indeed, assuming the validity of the parabolic 
approximation, the slope of the symmetry energy is proportional to the pressure of pure neutron matter 
at saturation density. That is,
\begin{equation}
  P_{{}_{\rm PNM}}(\rho_{0}) = \frac{1}{3}L\rho_{0}.
 \label{PNM}
\end {equation}

\subsection{Bayesian Refinement}
\label{Sec:Bayes}

The demand for robust quantification of uncertainties associated with model calculations of 
physical observables\,\cite{PhysRevA.83.040001} has motivated the calibration of a certain 
class of covariant EDFs\,\cite{Chen:2014sca,Chen:2014mza} that aim to describe both the
properties of finite nuclei and neutron stars. Initially, the model parameters 
were obtained from the minimization of a suitable objective (or ``cost") function constructed 
from binding energies, charge radii, and isoscalar-monopole excitations of a variety of
spherical nuclei. The minimization was then supplemented by a covariance analysis that 
explores the landscape around the minimum, thereby providing uncertainty estimates and 
correlation coefficients. Although accurately calibrated to a host of nuclear properties, the
various models described in Refs.\,\cite{Chen:2014sca,Chen:2014mza} differ---often
dramatically---in their predictions of observables the are highly sensitive to the density 
dependence of the symmetry energy. As such, our goal is to use covariance matrices
extracted from these earlier studies as the prior distribution of parameters to be
refined by the inclusion of a variety of neutron star observables that have been collected 
during the last five years. These include the tidal deformability of a 1.4\,$M_{\odot}$ neutron 
star extracted from GW170817\,\cite{Abbott:PRL2017,Abbott:2018exr}, stellar radii deduced 
from the NICER mission\,\cite{Riley:2019yda,Miller:2019cac,Miller:2021qha,Riley:2021pdl}, 
and limits on the most massive neutron stars obtained from pulsar timing 
observations\,\cite{Demorest:2010bx,Antoniadis:2013pzd,Cromartie:2019kug,Fonseca:2021wxt}. 
In addition, given the success of $\chi$EFT in reproducing low-energy properties of finite 
nuclei, we incorporate predictions on the behavior of pure neutron matter to inform the EOS 
at low densities.

For refining two existing models---FSUGold2 with a stiff symmetry energy\,\cite{Chen:2014sca} 
and FSUGarnet with a soft one\,\cite{Chen:2014mza}---we use the most optimistic estimates 
for the mass and radius of the two pulsars PSR J00740+6620 and PSR J0030+0451 targeted 
by the NICER mission\,\cite{Miller:2021qha}. Further, we also include the tidal deformability of a 
1.4\,$M_{\odot}$ neutron star recommended by the Ligo-Virgo collaboration\,\cite{Abbott:2018exr}. 
This new information is displayed in the following set of equations: 
\begin{subequations}
\begin{align}
\text{PSR J0740+6620} \quad \quad &R=12.35 \pm 0.75 \text{ km}\nonumber\\
&M=2.08 \pm 0.07 \ M_{\odot} \\
\text{PSR J0030+0451} \quad \quad &R=12.45 \pm 0.65 \text{ km}\nonumber\\
&M=1.44 \pm 0.15 \ M_{\odot} \\
\text{GW170817} \quad \quad & \Lambda_{1.4} = 190^{+390}_{-120}\label{GW170817}
\end{align} 
\label{NewData}
\end{subequations}
Note that the quoted errors are all 1$\sigma$ errors, with the exception of the tidal 
deformability that is quoted at the 90\% confidence level. Finally, we refine the low-density 
component of the EOS by incorporating theoretical predictions from 
$\chi$EFT\,\cite{Drischler:2021kxf}. 

To calibrate---or rather re-calibrate---the model parameters, we start by using the covariance matrices 
associated with the FSUGold2\,\cite{Chen:2014sca} and FSUGarnet\,\cite{Chen:2014mza} EDFs. 
These covariance matrices were obtained with a fitting protocol that used as input the binding energy, 
charge radius, and giant monopole resonance of a set of spherical nuclei as well as the maximum 
neutron star mass known at the time. In the present Bayesian refinement, such covariance matrices
will be used as the prior distribution of model parameters, which will then be combined with the new
data to generate the posterior distribution via Monte Carlo sampling.

Given the strong-coupling nature of the theory, it is ill-advised to change each model parameter individually, 
as the searching algorithm often ends up wandering aimlessly in the landscape of parameters. Moreover, 
the connection between the model parameters and our physical intuition is tenuous at best. In an effort to 
mitigate this problem we use a mapping between the model parameters and a few bulk properties of infinite 
nuclear matter that have a clear physical interpretation\,\cite{Chen:2014sca}. That is, the set of coupling 
constants that appear in Eq.(\ref{LDensity}) that will be adjusted in response to the new data are:
$\textbf{C} = \{g_{\rm s}, g_{\rm v}, g_{\rho}, \kappa, \lambda, \zeta, \Lambda_{\rm v} \}$. With the
exception of the quartic vector coupling constant $\zeta$ that controls the high-density behavior of
the EOS, the other six coupling constants can be directly mapped to the following set of bulk properties 
of infinite nuclear matter evaluated at saturation density\,\cite{Chen:2014sca}: 
$\boldsymbol{\theta}\!=\!\{\EoverA_{0}, \rho_{0}, M^{\star}, K, J, L\}$, where $M^{\star}$ is the 
effective nucleon mass at saturation density $\rho_{0}$ and the remaining bulk parameters were defined 
in Eq.(\ref{EandS}). The coupling constant $\zeta$ is left as a free parameter. Predictions from FSUGold2 
and FSUGarnet for the central values of this set of bulk properties are listed in Table\,\ref{Table1}. 

Evidently, it is more natural and intuitive to estimate a suitable range of values for the bulk parameters 
$\boldsymbol{\theta}$ than for the model parameters {\bf C}. For example, the simple liquid-drop model 
already provides good estimates for both $\EoverA_{0}$ and $J$. Further, because covariant EDFs 
are characterized by the presence of strong scalar and vector fields, changing one model parameter at 
a time hinders the convergence of the self-consistent procedure required to solve the mean-field equations. 
Instead, modifying a single bulk parameter does not sacrifice the convergence, as it involves a coherent 
change of several model parameters. Finally, we observe that whereas the bulk parameters associated 
with symmetric nuclear matter are narrowly constrained, the situation is drastically different for the symmetry 
energy,  especially in the case of its slope $L$. That the density dependence of the symmetry energy is poorly 
constrained is a limitation of the existing database of nuclear data that lacks observables with very large 
proton-neutron asymmetries.

\onecolumngrid
\begin{center}
\begin{table}[h]
\begin{tabular}{|m{2.1cm}|| m{2.1cm} m{2.1cm} m{2.1cm} m{2.1cm} m{2.1cm} m{2.1cm} m{2.1cm}|}
\hline\rule{0pt}{2.5ex}   
  \!Model & $\rho_0 (\text{fm}^{-3})$ & $\EoverA_{0}$ (MeV) & $M^{\star}/M$ & K (MeV) & J (MeV) & L (MeV) & $\zeta$ \\
    \hline\hline
    FSUGold2 & 0.1505 & -16.28 & 0.593 & 238.00 & 37.62 & 112.8 & 0.0256 \\
    FSUGarnet & 0.1531 & -16.23 & 0.578 & 229.62 & 30.92 & 50.96 & 0.0234 \\
    \hline \hline
\end{tabular}
\caption{Central values for various bulk properties of infinite nuclear matter as predicted by
              the two covariant EDFs used in this work: FSUGold2\,\cite{Chen:2014sca} and 
              FSUGarnet\,\cite{Chen:2014mza}.}
\label{Table1}
\end{table}
\end{center}
\twocolumngrid
Such unfavorable situation has changed dramatically by the recent measurements of neutron star properties that provide 
vital information on the symmetry energy around twice nuclear matter saturation density. In turn, $\chi$EFT predictions for
the EOS of pure neutron matter fill an important gap below saturation density. 

In the context of Bayes theorem, the new information provides valuable constraints for refining our model. In essence, 
Bayes theorem describes how to update our current knowledge (or ``belief") given some new evidence. In mathematical 
form, Bayes theorem is written as follows\,\cite{Gregory:2005,Stone:2013}:
\begin{equation} \label{bayes1}
 P(\mathcal{M}|D) = \dfrac{P(D|\mathcal{M}) P(\mathcal{M})}{P(D)}.
\end{equation}
Here $P(\mathcal{M})$ contains our prior knowledge of the model parameters before the new data is incorporated. 
This knowledge is summarized in the covariance matrix associated with earlier calibrations of the EDFs. For example, 
FSUGold2 predicts a stiff symmetry energy characterized by a slope of $L\!=\!(112.8\!\pm\!16.1)$\,MeV\,\cite{Chen:2014sca}. 
Given that  $\chi$EFT favors a relatively soft symmetry energy, our present knowledge of the symmetry energy will 
likely be updated as a result of this new evidence. The new information is incorporated into the conditional probability 
$P(D|\mathcal{M})$, often also referred to as the likelihood function $\mathcal{L}(\mathcal{M},D)$. Finally, $P(D)$ is 
a normalization factor known as the marginal probability. These three quantities are combined according to Eq.(\ref{bayes1}) 
to define the posterior distribution $P(\mathcal{M}|D)$, namely, the updated probability distribution that emerges from
the new evidence. In this work we will sample the posterior distribution of parameters by using a Markov Chain Monte 
Carlo (MCMC) method. As a result, the marginal distribution $P(D)$ plays no role since the MCMC method is only
sensitive to the ratio of probabilities. The posterior distribution of parameters may then be written as 
\begin{equation} \label{bayes2}
 P(\mathcal{M}|D) \propto \mathcal{L}(\mathcal{M},D) P(\mathcal{M}).
\end{equation}

Given our knowledge of the prior distributions $P(\mathcal{M})$ for both FSUGold2 and FSUGarnet, all the remains 
is to specify the structure of the likelihood function. To do so, we introduce an objective function $\chisq$ that is 
defined in terms of the sum of the squared residuals between the experimental observables and the associated 
theoretical predictions. That is, the likelihood function is defined as follows:
\begin{subequations}
\begin{align}
\mathcal{L}(\boldsymbol{\theta},D) & = \mathlarger{e}^{-\frac{1}{2}\chisq(\boldsymbol{\theta},D)}\\
 \chisq(\boldsymbol{\theta},D) & = \sum_{n=1}^{N}
 \frac{\Big(\mathcal{O}_{n}^{\rm (th)}(\boldsymbol{\theta})-\mathcal{O}_{n}^{\rm (exp)}\Big)^{2}}
 {\Delta\mathcal{O}_{n}^{2}} \,.
 \label{likelihood}
\end{align}
\end{subequations}
Incorporating into the $\chisq$ function the tidal deformability of a $1.4\,M_{\odot}$ neutron star quoted in Eq.(\ref{GW170817}) 
as well as the $\chi$EFT predictions for the EOS of pure neutron matter is straightforward. However, 
including mass and radius information from both NICER sources is slightly more complicated given that both observables are 
quoted with their own uncertainties. For these cases, the likelihood function involves a generalized two dimensional $\chisq$
function given by a line integral over the predicted mass-radius curve\,\cite{Frohner:1996,Steiner:2018rxu,Brandes:2022nxa}. 
That is, for a given parameter set $\boldsymbol{\theta}$, one generates a parametric curve of masses $M(\boldsymbol{\theta},s)$ 
and associated radii $R(\boldsymbol{\theta},s)$ parametrized in terms of a generic parameter $s$; for example, the central 
pressure. Assuming no correlation between the mass and radius measurements, the likelihood function for an observed 
neutron star with mass $M^{({\rm exp})}$, radius $R^{({\rm exp})}$, and associated errors $\sigma_{\rm M}$ and $\sigma_{\rm R}$, 
is given by a line integral of the following form:
\begin{widetext}
\begin{equation}
 \mathcal{L}\Big(\boldsymbol{\theta},M^{({\rm exp})},R^{({\rm exp})}\Big) \propto 
  \int_{\Gamma}
 \exp\!\left[-\frac{1}{2}\left(\frac{M(\boldsymbol{\theta},s)-M^{({\rm exp})}}{\sigma_{\rm M}}\right)^{\!\!2}\,\right]
 \exp\!\left[-\frac{1}{2}\left(\frac{R(\boldsymbol{\theta},s)-R^{({\rm exp})}}{\sigma_{\rm R}}\right)^{\!\!2}\,\right] ds,
 \label{Bivariate}
\end{equation}
\end{widetext}
where $ds$ represents the line element along the parametric mass-radius curve $\Gamma$. Note that if the stellar mass is known 
with arbitrary precision, i.e., $\sigma_{\rm M}\!\rightarrow\!0$, then the associated exponential becomes a Dirac delta function and 
the likelihood reduces to a standard univariate distribution in the stellar radius. Moreover, the likelihood function accounts for those 
cases in which the EOS associated with a certain set of parameters $\boldsymbol{\theta}$ may not be sufficiently stiff to support 
very heavy neutron stars, such as PSR J0740+6620. In such cases, the parameter set $\boldsymbol{\theta}$ is rejected with a 
probability proportional to Eq.(\ref{Bivariate}) rather than being flatly rejected. Finally, the likelihood function used in this work 
becomes a product of individual likelihood functions for the tidal deformability $\Lambda_{1.4}$, for masses and radii of the two 
NICER sources PSR J0740+6620 and PSR J0030+0451, and for the EOS of pure neutron matter as predicted by $\chi$EFT. 
Such a likelihood function $\mathcal{L}(\mathcal{M},D)$ when combined with the prior distribution of parameters $P(\mathcal{M})$ 
defines the posterior distribution $P(\mathcal{M}|D)$ of Eq.(\ref{bayes2}). The posterior distribution will be sampled using a MCMC 
method implemented via a traditional Metropolis-Hastings algorithm. Once such posterior distribution of parameters is generated, 
one can readily obtain averages, standard deviations, and correlations for all observables of interest. For example, the predicted 
radius of a neutron star with a measured mass $M$ and associated error $\sigma_{M}$ is computed by averaging over the $N$ 
samples of the posterior distribution of parameters. That is,
\begin{equation}
 \big\langle R(M)\big\rangle \!\propto\! \sum_{n=1}^{N}  
  \int_{M_{\rm min}}^{M_{\rm max}}\hspace{-5pt} R_{n}
 \exp\!\left[-\frac{1}{2}\left(\frac{M_{n}\!-\!M}{\sigma_{M}}\right)^{\!\!2}\,\right]dM_{n},
 \label{AvgR}
\end{equation}
where the integral varies over the $[M_{\rm min},M_{\rm max}$] interval and the masses $M_{n}$ in such interval are generated from 
the ${\rm n}_{\rm th}$ sample of the posterior distribution. Here $M_{\rm min}$ is a suitable lower limit for the integral and $M_{\rm max}$ 
is the maximum mass generated by the ${\rm n}_{\rm th}$ parameter set. Finally, $R_{n}\!=\!R(M_{n})$ is the predicted radius for a 
neutron star of mass $M_{n}$.

\section{Results}
\label{Sec:Results}

The main goal of this work so to assess the impact that recent observational and theoretical information have on improving our
knowledge of the equation of state. As mentioned earlier, for the Bayesian refinement we adopt as prior distribution of parameters 
the covariance matrices obtained from the calibration of two EDFs: FSUGold2\,\cite{Chen:2014sca} and 
FSUGarnet\,\cite{Chen:2014mza}. The adoption of these priors guarantees that ground-state properties of spherical nuclei,
such as binding energies and charge radii, are accurately reproduced. However, the lack of information on the properties of
very neutron-rich nuclei leaves the isovector sector poorly determined. To assess the impact of the new information on the
isovector sector, we implement the Bayesian refinement in two stages. In the first stage we exclude $\chi$EFT predictions 
from the likelihood function, thereby relying exclusively on astrophysical information. In the second stage we add $\chi$EFT 
constraints to the astrophysical data.

\subsection{Refining EDFs with Astrophysical Data}

\begin{center}
\begin{figure}[h]
\centering
\includegraphics[width=0.99\textwidth]{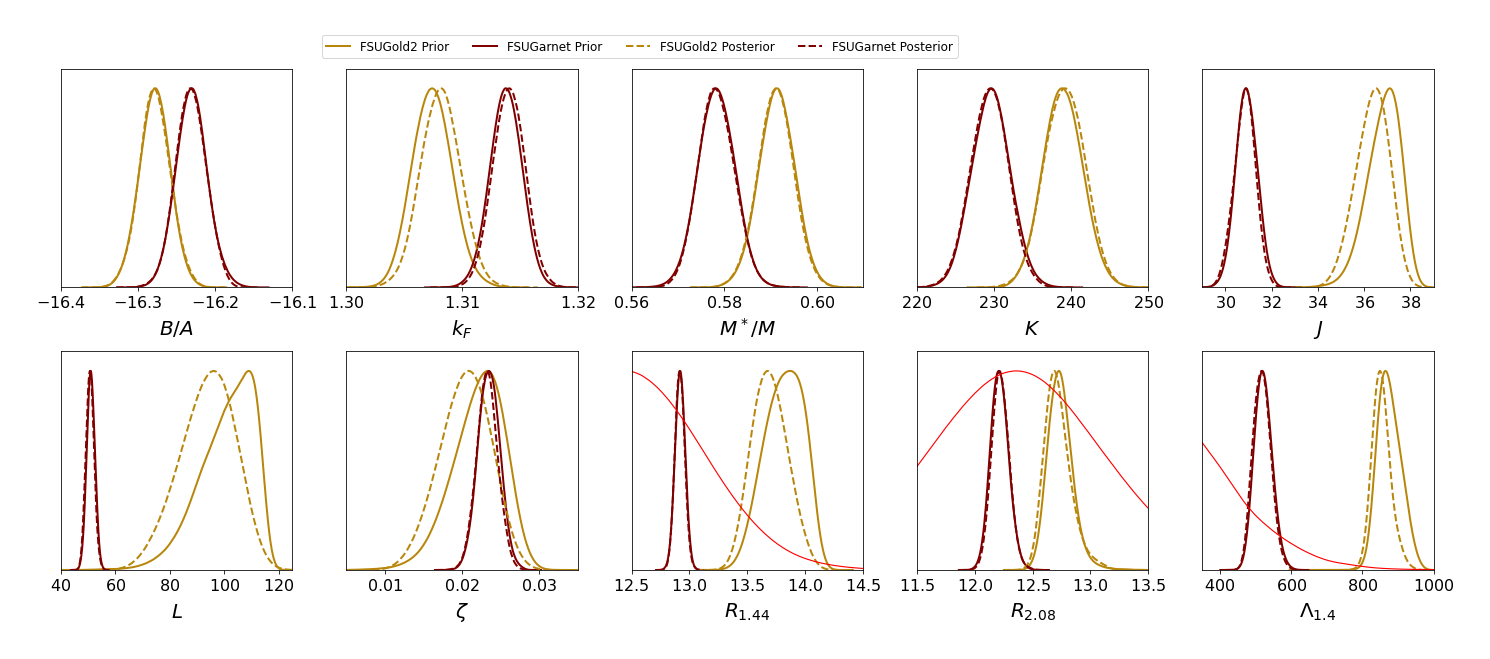}
\caption{MCMC results obtained by sampling the posterior distribution of parameters using only astrophysical data. The two 
covariant EDFs, FSUGold2 and FSUGarnet, are displayed with their respective colors. The solid red line depicts the probability 
distribution for the three astrophysical observables quoted in Eq.(\ref{NewData}). The binding energy per nucleon $B/A$, the 
incompressibility coefficient $K$, the symmetry energy $J$, and the slope of the symmetry energy $L$---all evaluated at saturation 
density---are given in MeV. The Fermi momentum at saturation $k_{\rm F}$ is given in units of fm$^{-1}$, stellar radii are reported 
in km, while the effective mass at saturation $M^{\ast}/M$, $\zeta$, and the tidal deformability $\Lambda_{1.4}$ are all dimensionless 
quantities. }
\label{Fig1}
\end{figure}
\end{center}

In this section we use the recent astrophysical information from both LIGO-Virgo and NICER collected in Eqs.(\ref{NewData}) 
to refine FSUGold2 and FSUGarnet. The impact of such a refinement is depicted in Fig.\ref{Fig1}, with the predictions obtained 
before the Bayesian refinement displayed by the solid lines while those after the refinement with dashed lines; FSUGold2 results
are displayed in gold whereas those from FSUGarnet in garnet. Also shown with the solid red line are the Gaussian probability
distributions for the three astrophysical observables quoted in Eq.(\ref{NewData}). As expected, there is practically no change 
post-refinement on the bulk properties of symmetric nuclear matter at saturation density, namely, on $B/A$, $k_{\rm F}$, 
$M^{\star}$, and $K$. This follows because both FSUGold2 and FSUGarnet were calibrated with nuclear observables that
are sensitive to the EOS of symmetric nuclear matter around saturation density. Moreover, we see hardly any changes on 
the predictions from FSUGarnet. Recall that FSUGarnet was calibrated assuming a soft symmetry energy, characterized by 
a slope of $L\!=\!50.96\,{\rm MeV}$, which seems to be favored by the astrophysical data. 

\onecolumngrid
\twocolumngrid

In contrast, the Bayesian refinement has a visible impact on various isovector quantities predicted by FSUGold2, such as 
$J$, $L$, $R_{1.44}$ and $\Lambda_{1.4}$. As opposed to FSUGarnet, FSUGold2 favors a fairly stiff symmetry energy. 
The astrophysical data seems to disfavor such a stiff symmetry energy and induces a mild softening. Such a softening of 
the symmetry energy must be compensated in order to be able to account for the existence of two-solar-mass neutron stars. 
Hence, the softening of the symmetry energy must be accompanied by a reduction of the coupling constant $\zeta$ that 
stiffens the EOS of symmetric nuclear matter in the high density region; see Fig.\ref{Fig1}.  Although some changes are 
clearly evident in Fig.\ref{Fig1}---especially in the case of FSUGold2---we conclude that for this particular set of covariant 
EDFs, astrophysical observations alone do not generate dramatic changes to the EOS, even after adopting the fairly 
optimistic errors in the neutron star radii quoted in Ref.\,\cite{Miller:2021qha}.

\subsection{Adding $\chi$EFT Information}

Whereas astrophysical data informs the EOS in the vicinity of two times saturation density, $\chi$EFT provides important
constraints at and below saturation density. For a model that predicts a fairly soft EOS such as FSUGarnet, we expect a
modest impact from $\chi$EFT. In contrast, we anticipate that the much stiffer FSUGold2 functional will be strongly 
affected by this new information. 
\onecolumngrid
\begin{center}
\begin{figure}[h]
\centering
\includegraphics[width=\textwidth]{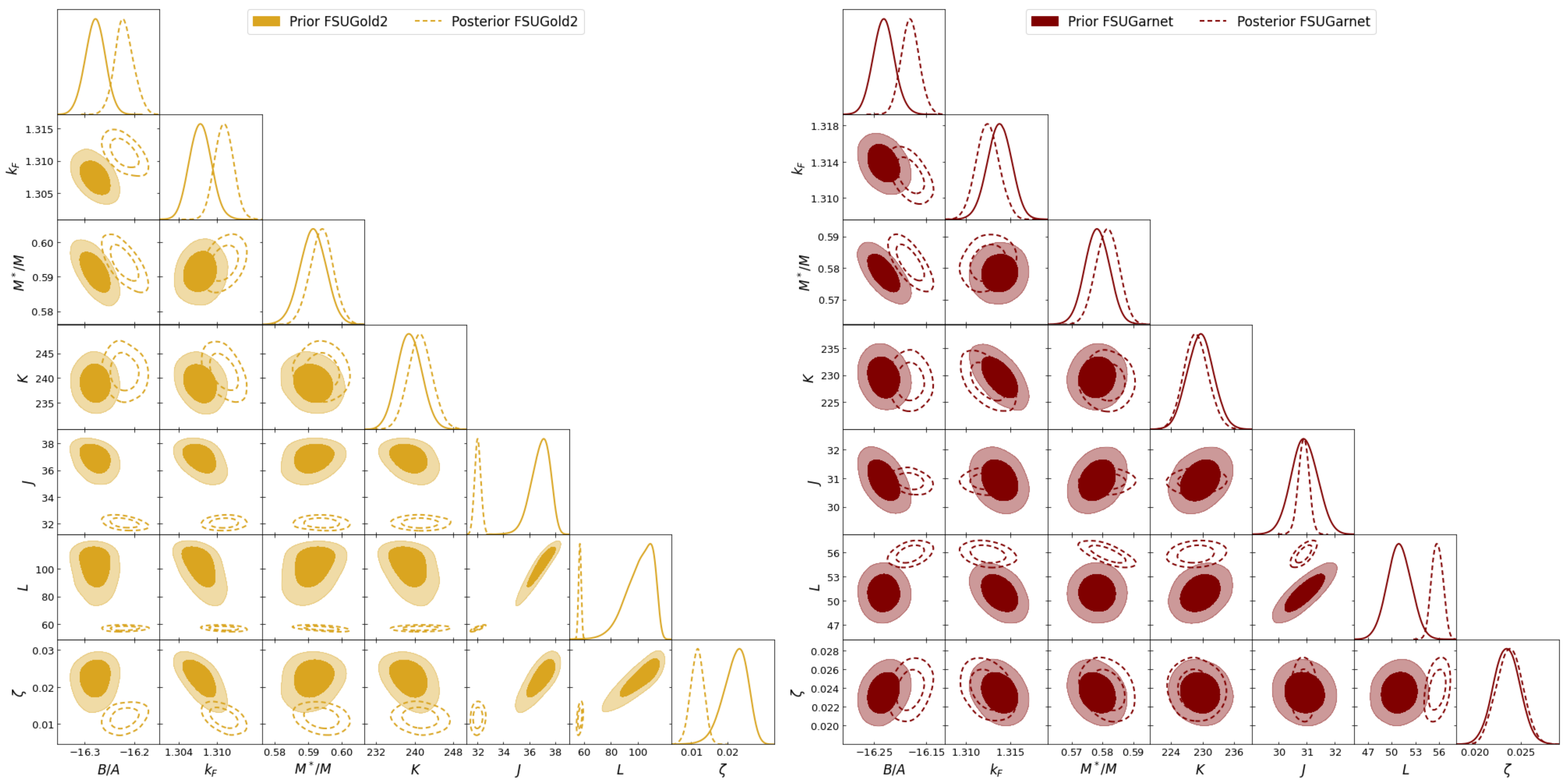}
\caption{Corner plot for both the prior and posterior distribution of bulk matter properties obtained after implementing
the Metropolis-Hasting algorithm. The posterior distributions (dashed lines) incorporate both astrophysical and $\chi$EFT 
constraints into the likelihood function. A total of 10 thousand MCMC steps were used to sample the posterior distribution 
and 50 thousand for the prior distribution. The ellipses represent 68\% and 95\% confidence intervals and we have adopted 
the same units and color convention as in Fig.\ref{Fig1}.}
\label{Fig2}
\end{figure}
\end{center}
\twocolumngrid
We display in Fig.\ref{Fig2} corner plots for both FSUGold2 and FSUGarnet obtained after implementing the 
Metropolis-Hasting algorithm. Unlike Fig.\ref{Fig1}, the posterior distribution of bulk parameters is informed
by a likelihood function that now contains both astrophysical and $\chi$EFT constraints. The covariance ellipses
displayed in the figure represent 68\% and 95\% confidence intervals. The first thing to notice is that $\chi$EFT 
shifts slightly the saturation point, an effect that is absent from Fig.\,\ref{Fig1} when only astrophysical information 
was used. Given that $\chi$EFT predicts a saturation point that is in conflict with the predictions from DFT, 
such a shift may have been expected. Relative to DFT predictions---which incorporate nuclear information 
into the calibration of the functional---$\chi$EFT tends to either saturate at higher density or to underbind 
the system\,\cite{Drischler:2021kxf}. However, as anticipated, the most dramatic changes involve the 
two symmetry energy parameters $J$ and $L$, especially for FSUGold2. Interestingly, in the case of FSUGarnet, 
$\chi$EFT actually favors a slight stiffening of the symmetry energy. Also noticeable is the significant reduction 
in the theoretical uncertainty in both $J$ and $L$, suggesting that the symmetry energy is much better constrained 
in $\chi$EFT than in DFT, where the calibration of the EDFs is hindered by the lack of isovector observables. This 
fact underscores the important role that high-order $\chi$EFT calculations play in the calibration of energy density 
functionals. We note that beyond the dramatic softening of the symmetry energy experienced by FSUGold2, a 
strong correlation also develops between $L$ and the isoscalar parameter $\zeta$. As already indicated, the 
quartic coupling $\zeta$ controls the high density component of the EOS, so if $L$ goes down, then $\zeta$ must 
compensate for such a change in order to be able to support $2{\rm M}_{\odot}$ neutron stars against gravitational 
collapse.

\onecolumngrid
\begin{center}
\begin{figure}[h]
\centering
\includegraphics[width=\textwidth]{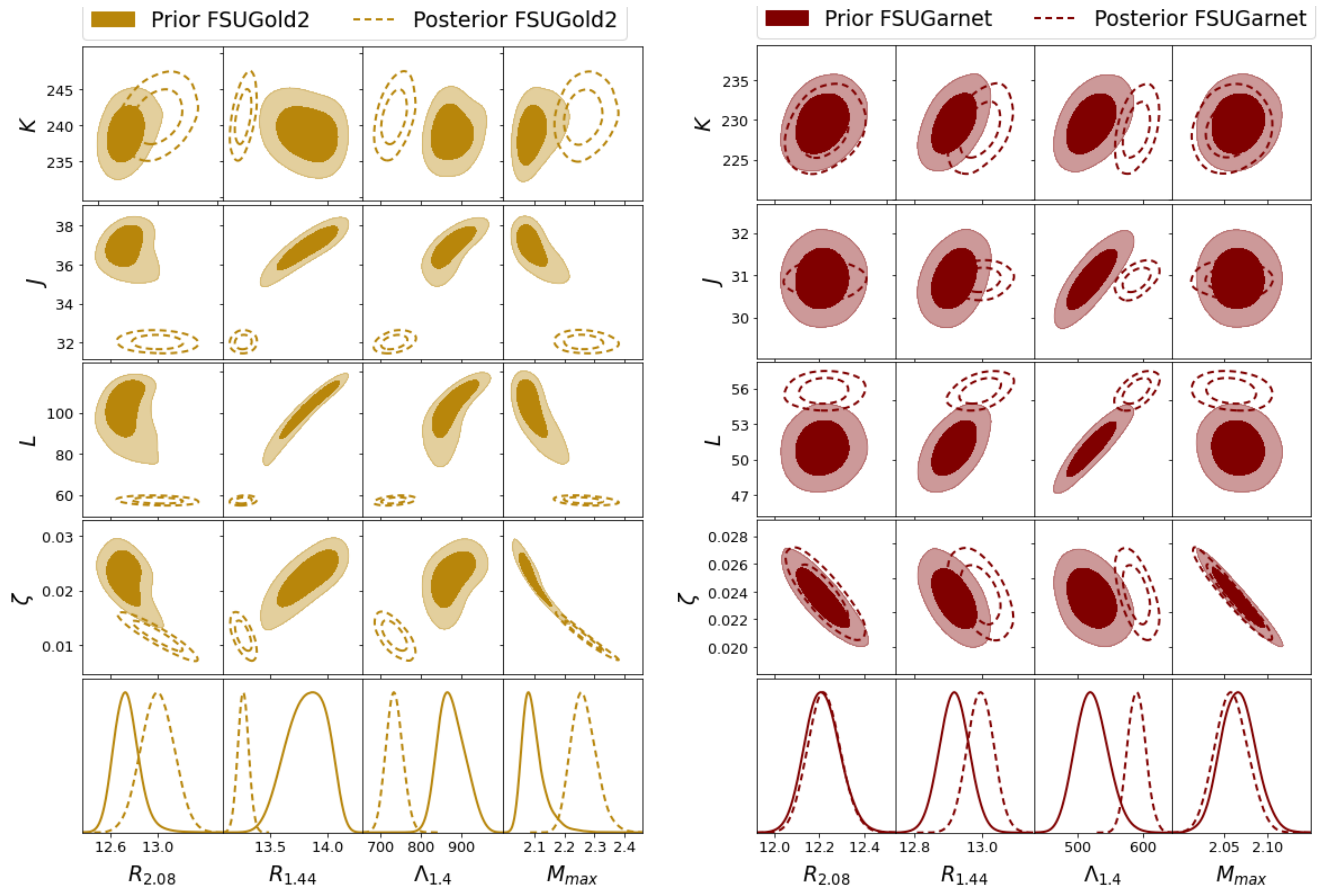}
\caption{Correlations between astrophysical observables and bulk properties after implementing
the Metropolis-Hasting algorithm. The posterior distributions (dashed lines) incorporate both astrophysical 
and $\chi$EFT constraints into the likelihood function. A total of 10 thousand MCMC steps were used to 
sample the posterior distribution and 50 thousand for the prior distribution. The ellipses represent 68\% 
and 95\% confidence intervals and we have adopted the same units and color convention as in Fig.\ref{Fig1}.}
\label{Fig3}
\end{figure}
\end{center}
\twocolumngrid

Having examined the statistical correlations between the various bulk parameters in Fig.\ref{Fig2}, we now proceed to
assess in Fig.\ref{Fig3} the impact of the Bayesian refinement on the astrophysical observables that were used to inform 
the likelihood function. As before, the impact of the refinement on FSUGarnet is a modest stiffening driven by $\chi$EFT 
that results in a slight increase to both the radius and tidal deformability of a $1.4{\rm M}_{\odot}$ neutron star. In the case 
of FSUGold2, it is interesting to note that the softening generated by $\chi$EFT reduces significantly the radius and tidal 
deformability of a $1.4{\rm M}_{\odot}$, but increases the maximum stellar mass and the radius of a $2{\rm M}_{\odot}$ 
neutron star. Again, this behavior is associated with the stiffening of the EOS at high densities required to compensate 
for the softening of the symmetry energy. Moreover, it also indicates---as expected---that the behavior of the symmetry 
energy at saturation density correlates poorly with the behavior of massive stars that is dominated by the EOS at densities 
that cannot be probed in terrestrial laboratories. 

\onecolumngrid
\begin{center}
\begin{figure}[h]
\centering
\includegraphics[width=0.8\textwidth]{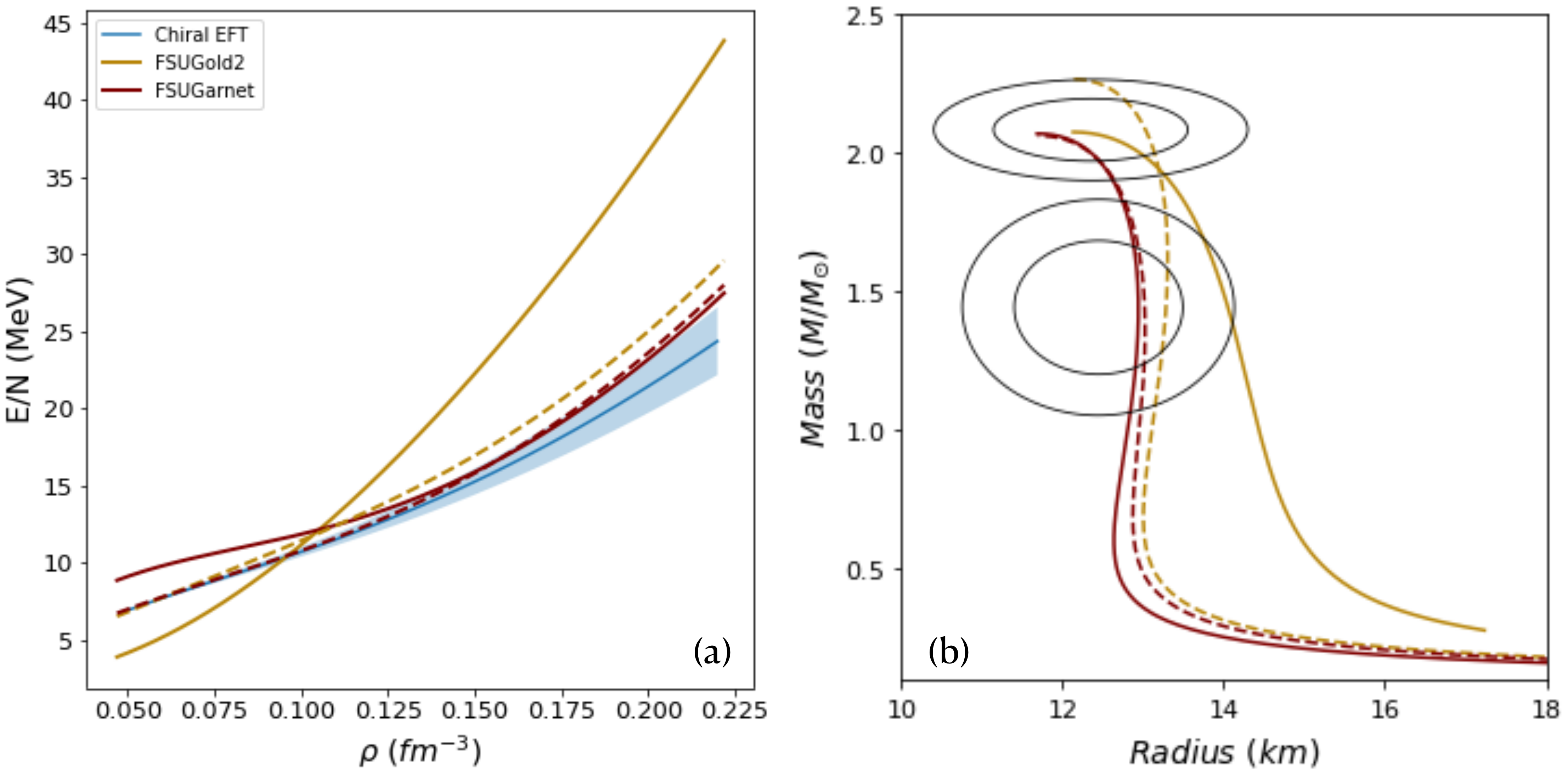}
\caption{(a) Equation of state for pure neutron matter as predicted by FSUGold2\,\cite{Chen:2014sca},
              FSUGarnet\,\cite{Chen:2014mza}, and a $\chi$EFT calculation correct up to 
              next-to-next-to-next-leading order (N${}^{3}$LO) in the chiral expansion\,\cite{Drischler:2021kxf}. 
              Predictions from FSUGold2 and FSUGarnet are displayed with their respective colors, with the solid
              and dashed lines representing results before and after the Bayesian refinement, respectively.
              (b) Mass-Radius relationship as predicted by FSUGold2 and FSUGarnet, both pre and post refinement
	      The covariance ellipses represent the 68\% and 95\% confidence intervals following Eq.(\ref{NewData}).}
\label{Fig4}
\end{figure}
\end{center}
\twocolumngrid

The impact of the refinement on the two covariant EDFs used in this work is summarized in 
Fig.\ref{Fig4}. Shown on the left-hand panel using the same color convention as in Fig.\ref{Fig1} are 
FSUGold2 and FSUGarnet predictions for the equation of state of pure neutron matter; for ease of viewing,
only central values are shown. In turn, the blue bands display $\chi$EFT predictions correct up to 
next-to-next-to-next-leading order\,\cite{Drischler:2021kxf}. As inferred from our previous discussion, the 
impact on FSUGarnet is modest, except at the lowest densities. The situation, however, is radically different 
in the case of FSUGold2. As we show below in Sec.\ref{Sec:Nskin}, the dramatic softening of the EOS will 
have important consequences on the prediction of the neutron skin thickness of ${}^{208}$Pb. Interestingly,
as shown on the right-hand panel in Fig.\ref{Fig4}, the softening induced at intermediate densities also 
generates a stiffening of FSUGold2 at the highest densities, resulting in a maximum neutron star mass of 
about $2.3{\rm M}_{\odot}$. Lastly, the covariance ellipses display in the figure represent the 68\% and 95\% 
confidence intervals for the two NICER measurements\,\cite{Miller:2021qha}. After refinement, the predictions 
of both FSUGold2 and FSUGarnet fall comfortably within the 68\% confidence ellipses.

The observation of neutron stars with masses in the vicinity of two solar masses requires a stiff EOS in 
order to support them against gravitational collapse\,\cite{Demorest:2010bx,Antoniadis:2013pzd,
Cromartie:2019kug,Fonseca:2021wxt}. Having incorporated both theoretical and observational
information into the model refinement, we now assess its impact on the EOS of neutron star matter 
$P\!=\!P(\Edens)$---and on the associated speed of sound defined as the derivative of the pressure
with respect to the energy density:
\begin{equation}
  \frac{c_{s}^{2}}{c^{2}} = \frac{dP(\Edens)}{d\Edens}.
 \label{SoundSpeed}
\end{equation}
Interest in the speed of sound in neutron stars was inspired by a conjecture grounded in holography 
that suggests that the conformal limit of $c_{s}^{2}/c^{2}\!=\!1/3$ represents an upper bound for a broad 
class of four-dimensional theories\,\cite{Cherman:2009tw}. In the context of neutron stars, Bedaque 
and Steiner suggested that the existence of heavy neutron stars is in strong tension with the conformal
limit\,\cite{Bedaque:2014sqa}. Later on, studies that incorporated the tidal polarizability as an additional
constraint on the EOS---both before\,\cite{Moustakidis:2016sab} and after\,\cite{Reed:2019ezm} 
GW170817\,\cite{Abbott:PRL2017}---also seem to suggest that the existence of heavy neutron stars 
is likely responsible for violating the conformal limit. Further, within the $\chi$EFT framework, it was 
found that the conformal limit is in tension with current nuclear physics constraints and observations 
of two-solar-mass neutron stars\,\cite{Tews:2018kmu}. The paper concludes with the fairly provocative 
statement that if the conformal limit holds at all densities, then nuclear physics models break down 
below twice saturation density.

\begin{center}
\begin{figure}[h]
\centering
\includegraphics[width=0.4\textwidth]{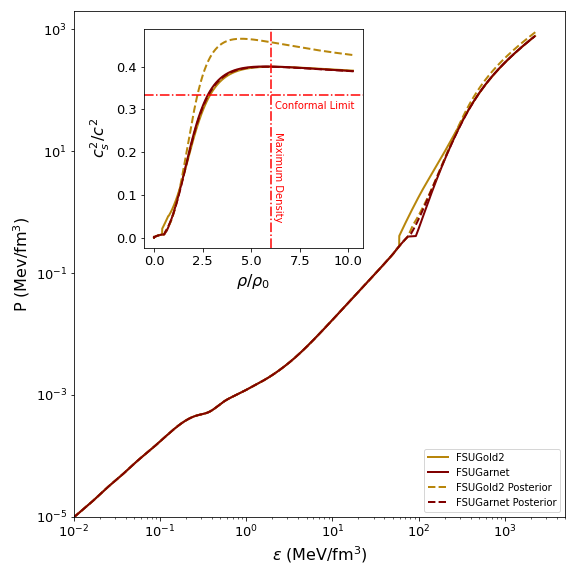}
\caption{Equation of state of neutron star matter and the associated speed of sound as predicted by
              FSUGold2 and FSUGarnet, with the solid and dashed lines representing results before and 
              after the Bayesian refinement, respectively. The red dot-dashed lines on the inset 
              indicate the conformal limit and the maximum density reached at the core at the maximum 
              mass configuration.}
\label{Fig5}
\end{figure}
\end{center}

To confront these assertions, we display in Fig.\ref{Fig5} predictions from both FSUGold2 and FSUGarnet 
before and after refinement. The EOS displays the various regions of the neutron star, particularly the 
outer crust, the inner crust, and the uniform liquid core. Concerning the speed of sound displayed on the 
inset in the figure, the conformal limit is already violated around two and a half times nuclear saturation 
density. Also shown on the inset is the maximum density reached at the center of the maximum mass 
configuration, a density that is significantly smaller than $\rho_{{}_{\!\rm pQCD}}\!\sim\!40\,\rho_{0}$, 
namely, the density at which QCD becomes perturbative and the conformal limit may be recovered. For 
a recent discussion on the role that perturbative QCD may play in constraining the EOS at neutron-star 
densities see Ref.\,\cite{Komoltsev:2021jzg} and references contained therein. 

\bigskip

We also observe on the inset of Fig.\ref{Fig5} that the the predictions of FSUGold2 after refinement suggest 
that the conformal limit is violated even earlier, at about twice $\rho_{0}$. This region of density is particularly 
interesting as it may be studied in the laboratory using energetic nuclear reactions that may compress and 
probe nuclear matter in the vicinity of twice nuclear saturation density. Indeed, this is one of the main science 
drivers behind the proposed FRIB-400 upgrade of the Facility for Rare Isotope Beams (FRIB). Such a rapid 
increase in the speed of sound displayed by the posterior distribution of FSUGold2 is driven 
by its prediction of a maximum neutron star mass of $2.3{\rm M}_{\odot}$; see Fig.\ref{Fig4}. Particularly 
relevant to this fact is the recent report of the ``black-widow" system PSR J0952-0607 with an extremely 
large pulsar mass of $(2.35\!\pm\!0.17){\rm M}_{\odot}$\,\cite{Romani:2022jhd}. If the existence of such 
massive neutron stars can be confirmed with better statistics, then---and not withstanding 
the small tidal deformability reported by the LIGO-Virgo collaboration\,\cite{Abbott:2018exr}---the 
violation of the conformal limit in neutron star interiors seems unavoidable.

Before leaving this section we list in Table\,\ref{Table2} the optimal parameter set for both FSUGold2 and
FSUGarnet after refinement. We underscore, however, that these are optimal (or central) values, as the 
EDFs after refinement are properly described by a statistical distribution of model parameters. Using this 
optimal set of parameters, we list in Table\,\ref{Table3} central values for the resulting bulk parameters after 
the Bayesian refinement. Most notably is the softening of the symmetry energy of FSUGold2, largely
induced by the inclusion of $\chi$EFT information; see Table\,\ref{Table1} for comparison.

\onecolumngrid
\begin{center}
\begin{table}[h]
\begin{tabular}{|l||c|c|c|c|c|c|c|c|c|c|}
\hline\rule{0pt}{2.5ex}   
\!Model   &  $m_{\rm s}$  &  $m_{\rm v}$  &  $m_{\rho}$  &  $g_{\rm s}^2$  &  $g_{\rm v}^2$  &  $g_{\rho}^2$  
                  &  $\kappa$       &  $\lambda$    &  $\zeta$       &   $\Lambda_{\rm v}$  \\
\hline
\hline
FSUGold2+R   & 501.611 & 782.500 & 763.000 & 103.760 & 169.410 & 128.301 & 3.79239  & $-$0.010635 & 0.011660 & 0.0316212  \\
FSUGarnet+R  & 495.633 & 782.500 & 763.000 & 109.130 & 186.481 & 142.966 & 3.25933  & $-$0.003285 & 0.023812 & 0.038274  \\
\hline
\end{tabular}
\caption{Central values for the model parameters FSUGold2 and FSUGarnet after Bayesian refinement. The parameter $\kappa$ 
and the meson masses $m_{\rm s}$, $m_{\rm v}$, and $m_{\rho}$ are all given in MeV, and the nucleon mass has been fixed at 
$M\!=\!939$ MeV.}
\label{Table2}
\end{table}
\end{center}
\twocolumngrid

\onecolumngrid
\begin{center}
\begin{table}[h]
\begin{tabular}{|m{2.1cm}|| m{2.1cm} m{2.1cm} m{2.1cm} m{2.1cm} m{2.1cm} m{2.1cm} m{2.1cm}|}
\hline\rule{0pt}{2.5ex}   
  \!Model  & $\rho_0 (\text{fm}^{-3})$ & $\EoverA_{0}$ (MeV) & $M^{\star}/M$ & K (MeV) & J (MeV) & L (MeV) & $\zeta$ \\
    \hline\hline
    FSUGold2+R  & 0.1522 & -16.22 & 0.594 & 241.22 & 32.03 & 57.20 & 0.0117 \\
    FSUGarnet+R & 0.1527 & -16.18 & 0.582 & 228.77 & 30.89 & 55.79 & 0.0238 \\
    \hline 
\end{tabular}
\caption{Central values for various bulk properties of infinite nuclear matter as predicted by
              FSUGold2 and FSUGarnet after the Bayesian refinement; compare to Table\,\ref{Table1}.}
\label{Table3}
\end{table}
\end{center}
\twocolumngrid
\vfill

\subsection{Heaven and Earth}
One of the most captivating features of neutron stars is the powerful connection between laboratory experiments
and astronomical observations. For example, it is well known that the slope of the symmetry energy $L$, which
can be determined by measuring the neutron skin thickness of heavy nuclei, controls the radius of low-mass 
neutron stars\,\cite{Horowitz:2000xj,Horowitz:2001ya,Carriere:2002bx,Lattimer:2006xb}. One expects, however, 
that the correlation between the thickness of the neutron skin and the neutron star radius will weaken as the stellar 
mass increases. It is the aim of this section to identify the density region that has the strongest impact on the 
development of the stellar radius.

To do so we implement the following procedure. First, as part of the Bayesian refinement, we generate multiple 
samples of various observables distributed according to the posterior distribution. Second, for each of the samples, 
we isolate the entire mass-radius relationship and the associated equation of state. We then proceed to store the 
pressure $P_{\!\rho}$ at various values of the baryon density $\rho$ as well as the predicted stellar radii $R_{M}$ over 
a given mass ($M$) range.  Finally, once all Monte Carlo samples have been generated, we compute the Pearson 
correlation coefficient between $R_{M}$ and $P_{\!\rho}$.

\onecolumngrid
\begin{center}
\begin{figure}[h]
\centering
\includegraphics[width=0.95\textwidth]{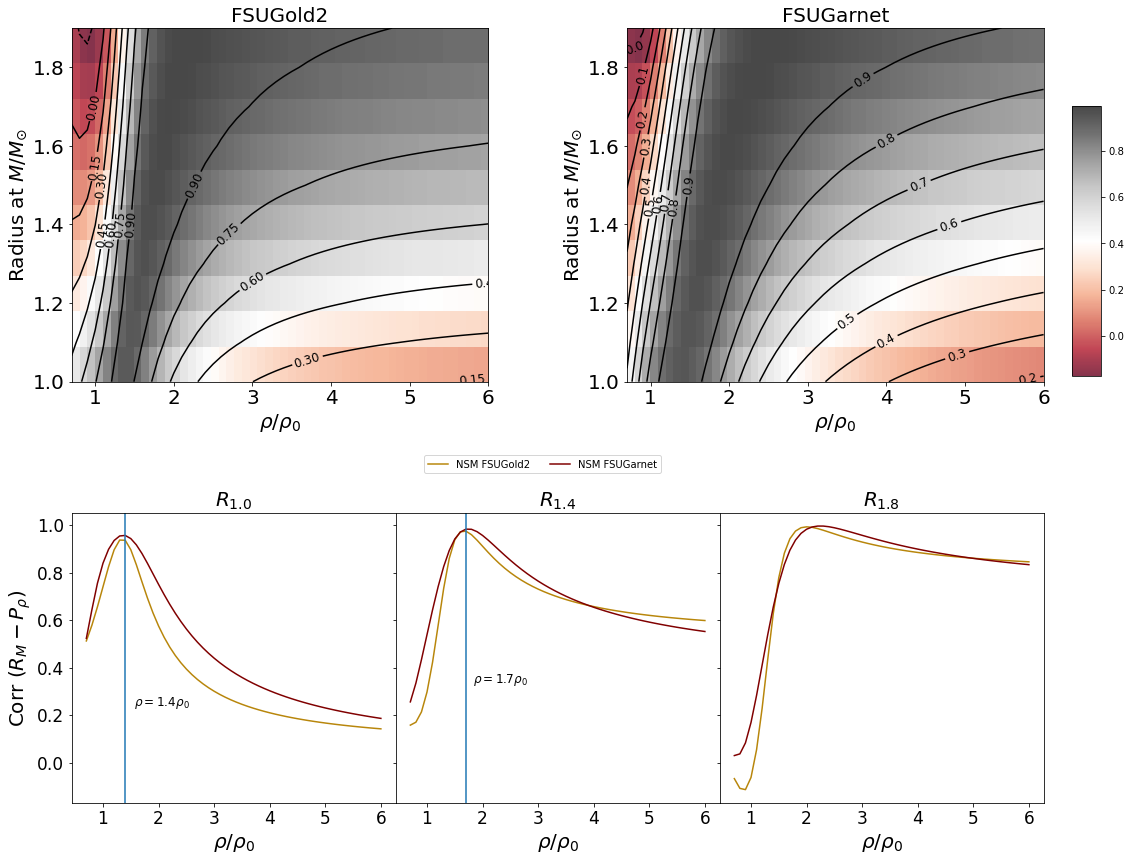}
\caption{Pearson correlation coefficients inferred from the posterior distribution of FSUGold2 and FSUGarnet. 
              The top panel displays contour plots labeled by the correlation coefficient between the stellar radius
              of a given mass configuration $R_{M}$ and the associated pressure support as a function of density  
	      $P_{\!\rho}$. The bottom panel displays the $R_{M}\!-\!P_{\!\rho}$ correlation coefficient for three
	      neutron stars with individual masses of 1.0, 1.4, and 1.8\,${\rm M}_{\odot}$.}
\label{Fig6}
\end{figure}
\end{center}
\twocolumngrid

Shown on the top panel of Fig.\,\ref{Fig6} is a heat map that displays the $R_{M}\!-\!P_{\!\rho}$ correlation predicted from
the FSUGold2 and FSUGarnet posterior distribution, with the various contour lines labeled according to the value of the 
correlation coefficient. For example, in the case of FSUGold2, the pressure in the narrow $(1.2\!-\!1.5)\,\rho_{0}$ density 
region correlates with the radius of a $1\,{\rm M}_{\odot}$ neutron star to better than 90\%. This behavior is better
appreciated in the lower panel of Fig.\,\ref{Fig6}, which clearly indicates that the radius of a $1\,{\rm M}_{\odot}$ neutron 
star is dominated by the pressure over a very narrow range of densities. This validates the claim that laboratory experiments
that determine the neutron skin thickness of heavy nuclei place stringent constraints on the radius of low-mass neutron 
stars\,\cite{Carriere:2002bx}. For a ``canonical'' $1.4\,{\rm M}_{\odot}$ neutron star, the strongest correlation develops at
$1.7\,\rho_{0}$, yet larger values of the density that are no longer accessible in the laboratory continue to make an 
important contribution. Finally, for a $1.8\,{\rm M}_{\odot}$ neutron star, the relevant region of densities is too high and wide 
for laboratory experiments to play a pivotal role in the determination of the stellar radius. 

\subsection{Neutron Skins}
\label{Sec:Nskin}

In the previous sections we have demonstrated how the Bayesian refinement of two previously calibrated 
covariant EDFs provides updated predictions that are in agreement with a large set of observables ranging 
from the properties of finite nuclei to the structure of neutron stars. Perhaps the only exception noticed so
far is the tidal deformability of a $1.4\,{\rm M}_{\odot}$ neutron star predicted by FSUGold2. Even after the
significant softening of the symmetry energy, FSUGold2 still predicts a tidal deformability of 
$\Lambda_{1.4}\!\approx\!740 \pm 40$ that lies significantly outside the 90\% confidence interval of 
$\Lambda_{1.4}\!=\!190^{+390}_{-120}$ quoted by the LIGO-Virgo collaboration\,\cite{Abbott:2018exr}. We 
note, however, that such a small tidal deformability is not without question. Indeed, a recent analysis that 
excluded waveform information beyond a certain frequency suggests significant larger values for the tidal 
deformability; see in particular Fig.8 of Ref.\,\cite{Gamba:2020wgg}. Under such a scenario, the refined 
FSUGold2 prediction for $\Lambda_{1.4}$ is no longer excluded. 

There is, however, a laboratory observable that seems to disfavor the softening induced on FSUGold2: 
the neutron skin thickness of ${}^{208}$Pb. Given the strong correlation between $L$ and the neutron skin 
thickness of ${}^{208}$Pb\,\cite{Brown:2000,Furnstahl:2001un,Centelles:2008vu,RocaMaza:2011pm}, we 
expect that the significant lower value of $L$ obtained after refinement will be in conflict with the PREX 
measurement\,\cite{Adhikari:2021phr}, which instead favors a fairly stiff symmetry energy\,\cite{Reed:2021nqk}.
Indeed, an interesting tension has emerged when confronting the PREX\,\cite{Adhikari:2021phr} and 
CREX\,\cite{Adhikari:2022kgg} measurements. Whereas the extracted neutron skin in ${}^{208}$Pb is
thick, CREX reported a very thin neutron skin in $^{48}$Ca; see Table\,\ref{Table4}. This presents a problem 
for the class of covariant EDFs used in this work, because the correlation between the neutron skins
of ${}^{208}$Pb and $^{48}$Ca is predicted to be strong\,\cite{Piekarewicz:2021jte,Giuliani:2022yna}. 

Besides the neutron skin thickness of ${}^{208}$Pb and $^{48}$Ca, we list in Table\,\ref{Table4} results for 
the charge and weak-charge radii, where the contribution from the finite nucleon size has been 
included\,\cite{Horowitz:2012we}. If the elastic form factor has been measured over a very wide range of 
momentum transfers, then the experimental radius may be extracted directly from the slope at the origin. 
This is the case for the charge radius of both $^{48}$Ca and ${}^{208}$Pb\,\cite{Angeli:2013}. Instead, given 
that PREX and CREX measured the weak form factor at only one momentum transfer, the extraction of the 
two weak-charge radii acquires a mild model dependence.

We observe in Table\,\ref{Table4} that the agreement between the FSUGold2 predictions and the PREX 
results are in excellent agreement, suggesting that the symmetry energy is indeed stiff\,\cite{Reed:2021nqk}. 
However, once the Bayesian refinement is implemented (FSUGold2+R) the excellent agreement is lost: the 
neutron skin thickness of ${}^{208}$Pb goes down from the experimentally consistent value o
f $R_{\rm skin}^{208}\!=\!0.285\,{\rm fm}$ all the way down to $R_{\rm skin}^{208}\!=\!0.203\,{\rm fm}$. 
In the case of $^{48}$Ca, the softening of the symmetry energy post-refinement moves the theoretical prediction 
in the direction of the experiment, but not nearly as much as it is required. That is, the neutron skin thickness of 
$^{48}$Ca, goes down from $R_{\rm skin}^{48}\!=\!0.231\,{\rm fm}$ to $R_{\rm skin}^{48}\!=\!0.197\,{\rm fm}$, 
which remains far from the quoted CREX value of $R_{\rm skin}^{48}\!=\!0.121(35)\,{\rm fm}$\,\cite{Adhikari:2022kgg}. 
It is important to note that after refinement the theoretical uncertainty in both $R_{\rm skin}^{208}$ and
$R_{\rm skin}^{48}$ is significantly reduced. 

\begin{center}
\begin{figure}[h]
\centering
\includegraphics[width=0.45\textwidth]{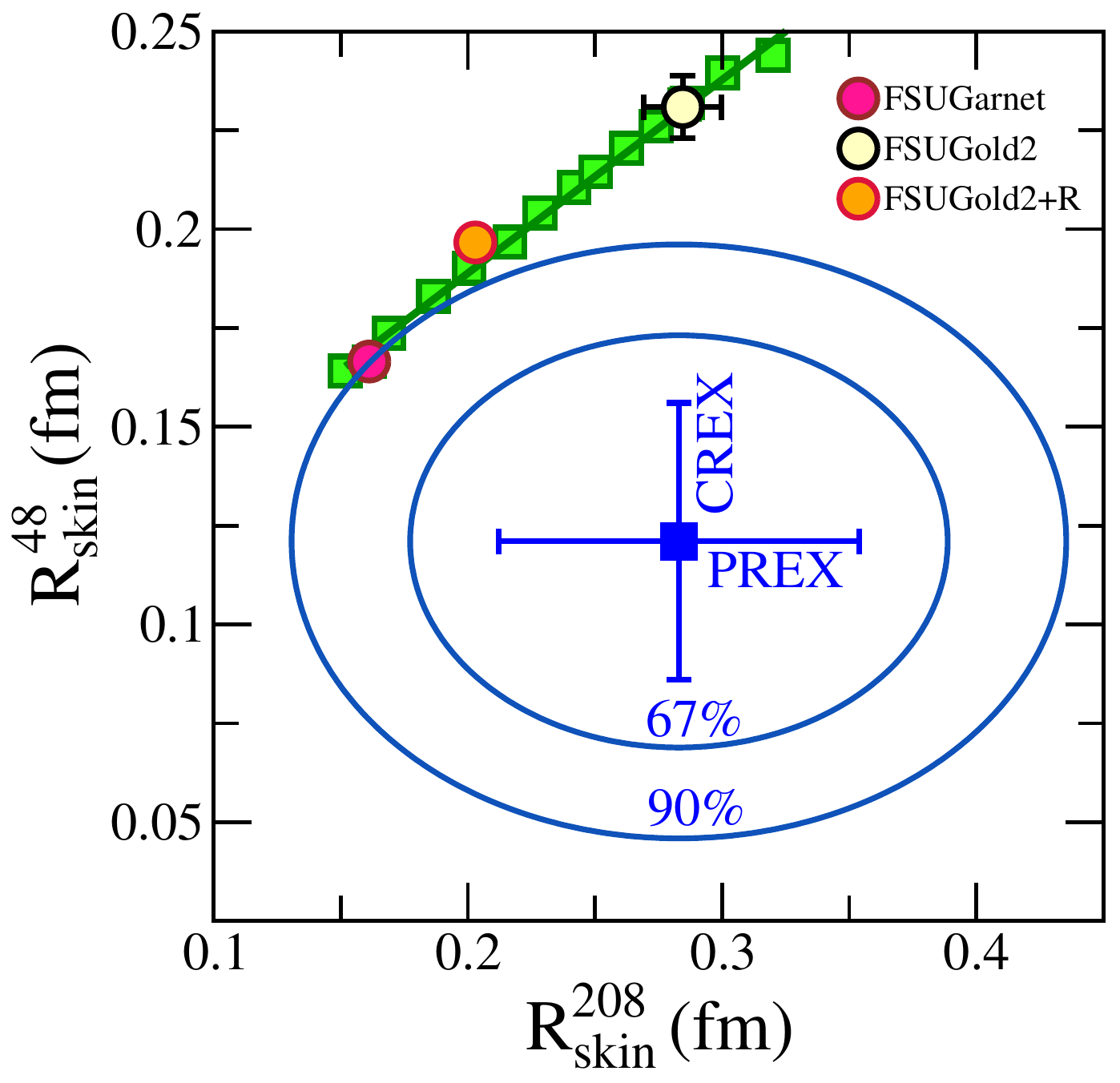}
\caption{Predictions for the neutron skin thickness of ${}^{208}$Pb and ${}^{48}$Ca from
              the same set of covariant EDFs used in Ref.\cite{Reed:2021nqk}, including the
              original FSUGarnet model. Also shown are predictions before and after refinement 
              from FSUGold2 and FSUGold2+R, respectively. Central values and 1$\sigma$ errors 
              together with 67\% and 90\% confidence ellipses are obtained from 
              Ref.\,\cite{Adhikari:2022kgg}.}
\label{Fig7}
\end{figure}
\end{center}

To further appreciate the present predicament, we display in Fig.\ref{Fig7}---alongside FSUGold2 and
FSUGold2+R---predictions for the neutron skin thickness of $^{48}$Ca and ${}^{208}$Pb using the 
same set of covariant EDFs employed in Ref.\,\cite{Reed:2021nqk} to analyze the PREX results. This
set includes the original (before refinement) FSUGarnet predictions. Also shown are the 67\% and 
90\% confidence ellipses together with the central values and 1$\sigma$ errors quoted by the 
PREX/CREX collaboration\,\cite{Adhikari:2022kgg}. The nearly perfect linear correlation between 
$R_{\rm skin}^{208}$ and $R_{\rm skin}^{48}$ is evident in the figure, yet none of the predictions reside 
inside the 90\% confidence ellipse. As indicated earlier, the original FSUGold2 prediction reproduces the 
PREX result, but grossly overestimates the CREX value. The Bayesian refinement does not improve the 
situation, as the new prediction simply slides along the regression line. 

\begin{table}[h]
\begin{center}
\begin{tabular}{|l||c|c|c|c|c|}
 \hline\rule{0pt}{2.25ex}   
 \!\!Model (${}^{208}$Pb)& $R_{\rm ch}$ & $R_{\rm wk}$ & $R_{\rm ch}\!-\!R_{\rm wk}$ & $R_{n}\!-\!R_{p}$ \\
 \hline
 \hline\rule{0pt}{2.25ex} 
 \!\!FSUGold2       &  5.491(6) &  5.801(19) & 0.310(16) & 0.285(15) \\ 
     FSUGold2+R  &  5.517(4) &  5.743(05) & 0.226(03) & 0.203(03)  \\ 
 \hline\rule{0pt}{2.25ex} 
     Experiment   &  5.501(1) &    5.800(75) & 0.299(75) & 0.283(71) \\
 \hline                                                                                                 
 \hline\rule{0pt}{2.25ex}   
 \!\!Model (${}^{48}$Ca)& $R_{\rm ch}$ & $R_{\rm wk}$ & $R_{\rm ch}\!-\!R_{\rm wk}$ & $R_{n}\!-\!R_{p}$ \\
 \hline
 \hline\rule{0pt}{2.25ex} 
 \!\!FSUGold2       &  3.426(3) &  3.707(07) & 0.281(08) & 0.231(08) \\ 
     FSUGold2+R  &  3.477(8) &  3.722(09) & 0.245(02) & 0.197(02)  \\ 
 \hline\rule{0pt}{2.25ex} 
     Experiment     &  3.477(2) &    3.636(35) & 0.159(35) & 0.121(35) \\
 \hline                                                                                                 
\end{tabular}
\caption{Predictions for FSUGold2 before and after refinement (+R) for the charge radius, weak radius,
              weak skin and neutron skin (all in fm) of ${}^{208}$Pb and ${}^{48}$Ca, as compared with the 
              experimental values extracted from PREX\,\cite{Adhikari:2021phr} and CREX\,\cite{Adhikari:2022kgg}.}
\label{Table4}
\end{center}
\end{table}

Perhaps the obvious deficiency displayed in Fig.\ref{Fig7} may be an indication that the physics encoded 
in the Lagrangian density given in Eq.(\ref{LDensity}) is incomplete. After all, with only two isovector 
parameters ($g_{\rho}$ and $\Lambda_{\rm v}$) it may be difficult to break the strong correlation between 
the neutron skins of $^{48}$Ca and ${}^{208}$Pb. We are currently working on extending the isovector sector 
of the covariant EDFs. Although models with a stronger theoretical underpinning may be able to reconcile 
both measurements at some level\,\cite{Hu:2021trw}, it is increasingly apparent that the skin-skin correlation 
can not be entirely broken. As such, one must conclude that---at present---no single theoretical framework can 
reproduce simultaneously the PREX and CREX results\,\cite{Reinhard:2022inh,Mondal:2022cva,
Yuksel:2022umn,Zhang:2022bni,Papakonstantinou:2022gkt}. 

Another possible resolution of the present dilemma is to appeal to the rather large experimental error bars. 
In this context, the only option is a more precise determination of $R_{\rm skin}^{208}$ at the future 
Mainz Energy-recovery Superconducting Accelerator (MESA)\,\cite{Becker:2018ggl}. Although a factor-of-two 
improvement in $R_{\rm skin}^{208}$ is realistic, the reality is that such an experiment is unlikely to be 
commissioned before the end of this decade. Regardless, we advocate for a more interesting resolution to
the dilemma, namely, that the answer is not hiding behind the experimental error bars but rather, that the 
answer may require to uncover some missing physics absent from existing theoretical descriptions.

\section{Conclusions}
\label{Sec:Conclusions}

Progress in our understanding of the equation of state of neutron star matter has grown significantly
during the last few years. The aim of this paper was to incorporate the latest information on neutron 
star properties to improve existing covariant energy density functionals that were largely calibrated by 
the properties of finite nuclei. Among the new properties informing the refinement of the functionals
are: maximum stellar masses reported from pulsar timing measurements\,\cite{Cromartie:2019kug,
Fonseca:2021wxt}, the simultaneous extraction of stellar radii and masses of two sources by the NICER 
mission\,\cite{Riley:2019yda,Miller:2019cac,Miller:2021qha,Riley:2021pdl}, tidal information from the 
LIGO-Virgo collaboration\,\cite{Abbott:PRL2017,Abbott:2018exr}, and predictions for the EOS of pure 
neutron matter from chiral effective field theory\,\cite{Drischler:2021kxf}. This new information was 
incorporated in a likelihood function which, through Bayesian inference, was used to refine two existing 
density functionals whose covariance matrices served as prior distributions of model parameters.

The two existing covariant EDFs, FSUGold2\,\cite{Chen:2014sca} and FSUGarnet\,\cite{Chen:2014mza}, 
followed a very similar fitting protocol to calibrated the model parameters. Given that the spherical 
nuclei used in the calibration are either stable or long-lived, both models predict similar isoscalar 
observables, such as the bulk properties of symmetric nuclear matter listed in Table\,\ref{Table1}. In 
contrast, at the time of the calibration, the database of strong isovector observables was very sparse, 
leading to a poorly constrained density dependence of the symmetry energy. Hence, the two EDFs 
adopted in this work were calibrated by selecting symmetry energies with a density dependence at 
the opposite ends of the spectrum, with FSUGold2 being stiff and FSUGarnet being soft. Motivated 
by the recent proliferation of strong isovector indicators, we implemented a Bayesian refinement of 
the model parameters by sampling posterior distributions via a Metropolis-Hastings algorithm.  

The refinement was implemented in two stages. We started by assessing solely the impact of 
astrophysical observables and later added $\chi$EFT predictions. When only astrophysical 
information was included, changes to the predictions of both models were modest, especially 
in the case FSUGarnet. For FSUGold2, the refinement slightly softens the symmetry energy 
resulting in mild reductions in both the slope of the symmetry energy $L$ and the radius of a 
$1.44\,{\rm M}_{\odot}$ neutron star. However, the situation changed dramatically once 
$\chi$-EFT predictions were incorporated. This we attribute to the sharp $\chi$EFT 
predictions relative to the fairly large astrophysical uncertainties. Although $\chi$EFT 
predicts a rather soft EOS for pure neutron matter, $\chi$EFT information stiffens the even 
softer symmetry energy predicted by FSUGarnet. However, the impact on FSUGold2 is 
dramatic: the symmetry energy at saturation density $J$, its slope $L$, and the non-linear
coupling $\zeta$ were all greatly sharpened and reduced; see Fig.\ref{Fig2}. Note that 
the reduction in the value of the non-linear coupling $\zeta$, which implies a stiffening of the
EOS of symmetric nuclear matter, is required to compensate for the softening of the 
symmetry energy. Without such a stiffening, the resulting EOS would not be able to support 
massive neutron stars. Indeed, after refinement, FSUGold2 predicts a maximum neutron star 
mass of about $2.3\,{\rm M}_{\odot}$, consistent with the black-widow pulsar with a 
mass of $(2.35\!\pm\!0.17){\rm M}_{\odot}$\,\cite{Romani:2022jhd}. As expected, the
induced softening of the symmetry energy reduces the stellar radius and tidal deformability
of a $1.4\,{\rm M}_{\odot}$ neutron star. However, the reduction in $\Lambda_{1.4}$ is not
sufficient to reproduce the fairly small value quoted by the LIGO-Virgo 
collaboration\,\cite{Abbott:2018exr}, although larger values do not seem to be completely
ruled out\,\cite{Gamba:2020wgg}.

Once both models were refined, we reported predictions for the mass-radius relation, the 
underlying EOS, and the associated speed of sound. Both of the predicted mass-radius
relations are fully consistent with the reported NICER values for both PSR J0740+6620 
and PSR J0030+0451. In particular, the existence of two-solar-mass neutron stars demands 
that the EOS be stiff at the highest densities encountered in the stellar core, which for the 
models under consideration reaches a value in the vicinity of $6\rho_{0}$. Notably, demands 
for a stiff EOS resulted in a violation of the conformal limit on the speed of sound at the relatively 
low densities of about $(2\!-\!2.5)\,\rho_{0}$, a range of densities that could be probed in the 
laboratory if the proposed FRIB-400 upgrade becomes a reality.

Whereas the Bayesian refinement of existing EDFs reproduce a large body of experimental and 
observational data over a broad range of densities, the induced softening of the symmetry 
energy is in conflict with recent results by the PREX and CREX collaborations on the neutron
skin thickness of ${}^{208}$Pb\,\cite{Adhikari:2021phr} and $^{48}$Ca\,\cite{Adhikari:2022kgg},
respectively. Although the experimental error bars are large, none of our models are able to
reconcile the very thick neutron skin in ${}^{208}$Pb with the very thin neutron skin in ${}^{48}$Ca.
From the perspective of covariant EDFs, the problem is especially challenging given that the neutron 
skin thickness of ${}^{208}$Pb and ${}^{48}$Ca are strongly correlated\,\cite{Piekarewicz:2021jte}.
Plans are currently under way to enlarge the isovector sector of the covariant EDFs in the hope 
of breaking such a correlation. However, the challenge to reconcile both measurements appears 
to go beyond the class of covariant EDFs considered in this work\,\cite{Reinhard:2022inh,
Mondal:2022cva,Yuksel:2022umn,Zhang:2022bni,Papakonstantinou:2022gkt}. Only time will 
tell whether the resolution of this puzzle is hiding behind the large experimental error bars or 
whether it demands a more ingenious solution.

\begin{acknowledgments}\vspace{-10pt}
We are grateful to Prof. Christian Drischler who provided $\chi$EFT predictions for the EOS 
of pure neutron matter. This material is based upon work supported by the U.S. Department 
of Energy Office of Science, Office of Nuclear Physics under Award Number DE-FG02-92ER40750. 
\end{acknowledgments}

\bibliography{./main.bbl}
\end{document}